\documentclass{desyproc}
\usepackage{aas_macros}
\newcommand{\jcap}{JCAP}

\begin{document}
\title{Searches for Axionlike Particles Using $\gamma$-Ray
Observations}

\author{{\slshape  Manuel Meyer on behalf of the \emph{Fermi}-LAT Collaboration}\\[1ex]
Department of Physics, Stockholm University, AlbaNova, SE-106 91 Stockholm, Sweden}

\contribID{familyname\_firstname}

\confID{13889}  
\desyproc{DESY-PROC-2016-XX}
\acronym{Patras 2016} 
\doi  

\maketitle

\begin{abstract}
Axionlike particles (ALPs) are a common prediction of theories beyond the Standard Model of particle physics that could explain the entirety of the cold dark matter. These particles could be detected through their mixing with photons 
in external electromagnetic fields. 
Here, we provide a short review over ALP searches that 
utilize astrophysical $\gamma$-ray observations. We summarize current 
bounds as well as future sensitivities and discuss the possibility that ALPs alter the $\gamma$-ray opacity of the Universe.
\end{abstract}

\section{Introduction}
Astrophysical observations with space-borne and ground based $\gamma$-ray experiments
have proven to be a powerful  tool to search for physics beyond the Standard Model. 
The currently operating \emph{Fermi} Large Area Telescope (LAT), sensitive to 
$\gamma$ rays between $20\,$MeV and above 300\,GeV \cite{atwood2009}, and
imaging air Cherenkov telescopes (IACTs), such as H.E.S.S., MAGIC, and 
VERITAS (sensitive to $\gamma$ rays above $\sim50$\,GeV \cite{hess,magic,veritas}) 
have provided strong constraints on, e.g.,  
Lorentz invariance violation \cite{Aharonian:2008kz,Ackermann:2009aa}, as well as on the annihilation cross section of weakly interacting massive particles,  which are prime cold dark matter candidates \cite{Ackermann:2015zua,hess:2016jja,Zitzer:2015eqa}.

Observations at $\gamma$-ray energies
can also be used to search for traces of axionlike particles (ALPs).
ALPs are closely related to the axion, which plays an essential role in the solution of the strong CP problem in QCD \cite{pq1977,weinberg1978,wilczek1978,jaeckel2010}.
Just as the axion, ALPs are well motivated cold dark-matter candidates \cite{preskill1983,abbott1983,dine1983,marsh2011,arias2012} that could be detected through their coupling to photons in external magnetic fields \cite{sikivie1983,raffelt1988}. They arise in several extensions of the Standard Model~\cite{witten1984,arvanitaki2010,ringwald2014review}.

Here, we review advancements in astrophysical $\gamma$-ray searches for ALPs that have, for certain ALP masses ($m_a$) below $\mu\mathrm{eV}$, reached a sensitivity comparable to that of current or future dedicated laboratory experiments. 
We give a short overview of the relevant astrophysical magnetic 
fields (Sec. \ref{sec:bfield}),\footnote{This review focuses on photon-ALP mixing in external magnetic fields, however, searches for $\gamma$-ray signals from axion and ALP decays have also been carried out \cite{Giannotti:2010ty,Berenji:2016jji}.} 
 and review the status of evidence found for 
a reduced $\gamma$-ray opacity that might be caused 
by photon-ALP oscillations (Sec. \ref{sec:opacity}). 
We review searches for ALP-induced spectral irregularities (Sec. \ref{sec:irreg})
before closing with an outlook on future observations (Sec. \ref{sec:out}).  
 
\section{Photon-ALP mixing at $\gamma$-ray energies}
\label{sec:bfield}
Solving the full equations of motion for the photon-ALP system, 
one realizes that the conversion probability
becomes maximal and independent of energy above 
a critical energy $E_\mathrm{crit} = |m_a^2 - \omega_\mathrm{pl}^2| / 2 g_{a\gamma} B$, where $B$ denotes the 
field strength transversal to the photon propagation direction, $g_{a\gamma}$
the photon-ALP coupling,
and $\omega_\mathrm{pl}$ the plasma frequency of the medium.
This strong mixing regime persists as long as $E < E_\mathrm{max} = 90\pi g_{a\gamma} B_\mathrm{cr}^2 / 7 \alpha B$, with $\alpha$ the 
fine structure constant and the critical magnetic field $B_\mathrm{cr}\sim4.4\times10^{13}$\,G. 
For photons around PeV energies, background photon fields such as the cosmic microwave background cause an additional photon dispersion leading to a modification of $E_\mathrm{max}$~\cite{dobrynina2015}.
Reviews on solving the equations of motion and
 deriving the photon-ALP mixing 
matrices in a number of magnetic fields are provided in, e.g., Refs. \cite{grossmann2002,csaki2003,mirizzi2008,bassan2010,deangelis2011,meyer2014}. 

One abundantly studied source class to search for traces of ALPs  
are blazars, active galactic nuclei (AGNs) with their 
relativistic jets closely aligned to the line of sight \cite{urry1995}.
Blazars make up the majority of extragalactic sources detected 
at $\gamma$-ray energies \cite{3fgl}. 
Their intrinsic brightness, especially during flaring episodes, 
and the numerous magnetic-field environments traversed 
by the photon beam along the line of sight 
make them excellent targets to search for ALP signatures.  
The photon-ALP oscillations may lead to two observables in the energy spectra of these objects (discussed in the following subsections):
a) the $\gamma$-ray source flux can be attenuated due to pair production with low energy photons originating from background radiation fields \cite{nikishov1962,gould1967,gould1967a}. ALPs produced in the vicinity of the source would circumvent this attenuation, and, if they reconvert to $\gamma$ rays, could lead to a significant boost of the observed photon flux, 
and b) oscillations of the flux should be imprinted in the spectra around  $E_\mathrm{crit}$ and $E_\mathrm{max}$ that depend on the morphology of the traversed $B$ fields. 

To accurately model the effect of ALPs on $\gamma$-ray spectra,
 a thorough understanding
of the intervening $B$ fields is necessary. 
Several different $B$-field environments have been studied in connection to photon-ALP conversions at $\gamma$-ray energies. 
They cover both coherent and turbulent magnetic fields, 
where the turbulent fields are often modeled 
with a simplified cell-like structure: each cell has a length equal to the coherence length $\lambda$ and the magnetic-field orientation changes randomly from one cell to the next. 
Starting from the blazar, the considered $B$ fields include: 
 the $B$-field in the AGN jet, 
\cite{sanchezconde2009,tavecchio2012,mena2013,tavecchio2014,meyer2014}, 
 the turbulent magnetic fields of the host galaxies
(usually found to be elliptical galaxies) that, because of the short 
coherence legnth of $\lambda \sim 0.1\,$kpc should not contribute significantly to photon-ALP mixing \cite{tavecchio2012,meyer2014},
 the $B$ field of the lobes of AGN jets 
\cite{tavecchio2014,meyer2014cta},
the turbulent $B$ fields of 
galaxy clusters that might host the blazar \cite{horns2012ICM,hess2013:alps,meyer2014}, 
 the intergalactic magnetic field (IGMF)\footnote{It was recently noted that photon-ALP mixing in the IGMF can be suppressed due to the photon dispersion for $\gamma$ rays above TeV energies \cite{Kartavtsev:2016doq}. 
}~\cite{deangelis2007,mirizzi2007,simet2008,mirizzi2009,sanchezconde2009,
 bassan2010, deangelis2011}
(for which only upper limits exist, $B \leqslant 1.7$\,nG  for $\lambda$ equal to the Jeans' length~\cite{pshirkov2016}),
and eventually the Galactic magnetic field (GMF) of the Milky Way \cite{simet2008,horns2012ICM}, which consists of 
both a turbulent and coherent component. Several models for the GMF have been put forward in the literature \cite{pshirkov2011,jansson2012}, with the best-fit values for the 
model of Ref. \cite{jansson2012} recently updated with measurements of the Planck satellite \cite{adam2016}.  

\subsection{Evidence for a reduced $\gamma$-ray opacity?}
\label{sec:opacity}
One predominant radiation field responsible for the
attenuation of  $\gamma$ rays originating from blazars 
with energies $10\,\mathrm{GeV}\lesssim E\lesssim 50\,\mathrm{TeV}$ is the extragalactic background light 
(EBL), which stretches from UV to far infrared wavelengths \cite{dwek2013}. The isotropic EBL photon density is difficult to measure directly
due to strong foreground contamination with zodiacal light \cite{hauser1998}. 
The EBL is composed of the emitted starlight integrated 
over the history of the Universe and the starlight that 
has been absorbed and re-emitted by dust in galaxies \cite{hauser2001,kashlinsky2005}.
The exponential attenuation scales with the optical depth $\tau(E,z)$,
a monotonically increasing function with $\gamma$-ray energy $E$, source redshift $z$, and EBL photon density (see Refs. \cite{franceschini2008,finke2010,kneiske2010,dominguez2011,gilmore2012,inoue2013} for a selection of EBL models). 

Using published IACT spectral data points, 
several authors have indeed found indications that state-of-the-art
EBL models over-predict the observed $\gamma$-ray attenuation. 
Once the observed spectrum has been 
corrected for the absorption utilizing an EBL model, 
such an over-prediction would manifest itself 
by a spectral hardening at energies corresponding
to a high optical depth.  
Accordingly, one way to search for such a feature 
is to fit power laws, $\phi(E) = dN/dE \sim E^{-\Gamma}$,
 with spectral 
indices $\Gamma_\mathrm{low}$ and $\Gamma_\mathrm{high}$ 
separately to the low and high energy part of the de-absorbed spectra.
If the opacity is over-predicted, 
the difference between the spectral indices $\Delta\Gamma = 
\Gamma_\mathrm{low} - \Gamma_\mathrm{high}$  
should increase with redshift as the predicted attenuation
 increases (this should also holds if the sources get dimmer 
 at higher energies). 
The above effect has been found 
and ALPs have been suggested as a possible solution \cite{deangelis2007,simet2008,deangelis2009,sanchezconde2009,
dominguez2011alps,deangelis2011,essey2012,rubtsov2014,galanti2015}.
Alternatively, it is possible to search for 
the spectral hardening by examining   
fit residuals $\chi = (\phi_\mathrm{obs} - \phi(E)e^{-\tau(E)})/\sigma_\phi$ of fits of smooth concave, i.e. non-hardening, functions $\phi(E)$ multiplied with EBL absorption to the observed spectra $\phi_\mathrm{obs}$ with flux uncertainties $\sigma_\phi$. 
Again, if $\tau$ is overestimated, the fit residuals should 
display a positive correlation with optical depth. 
A $4\,\sigma$ indication for such an effect was found \cite{horns2012,meyer2012ppa}
 that could be reduced 
significantly when photon-ALP oscillations are taken into account \cite{meyer2013}.
Intrinsic source effects leading to a spectral hardening
can fake such a signal (see e.g. Ref. \cite{lefa2011b} for one possibility),
 however, it would be 
highly contrived if such a hardening
occurred in several sources at 
exactly the energy where $\tau > 1$.

So far we have summarized observations of spectral hardening at high optical depths
that might hint at photon-ALP oscillations.
However, recent analyses could not confirm these
results. Using the largest IACT data set to date, 
the authors of Ref. \cite{biteau2015} repeated the analysis of Ref. \cite{horns2012}
and found no deviation from EBL-only expectations.
Furthermore, when including uncertainties
on the IACT energy resolution, and 
further systematic uncertainties, no spectral 
upturn can be observed when comparing \emph{Fermi}-LAT
and IACT spectra (spectra measured with the \emph{Fermi}-LAT 
should be less affected by EBL absorption at low redshift due to 
the lower energy range accessible with the satellite) \cite{Sanchez:2013lla}.  
Using a sample of \emph{Fermi}-LAT detected blazars above 50\,GeV and 
up to redshifts of $z \sim 2$,
again, no evidence for a spectral hardening was found 
in \emph{Fermi}-LAT data \cite{dominguez2015}.
It should be noted that the \emph{Fermi}-LAT 
has only detected  $\gamma$ rays up to $\tau \sim 3$ \cite{2fhl}, whereas the ALP effect should become particularly pronounced for $\tau \gtrsim 4$ \cite{meyer2014cta}.   

Future dedicated analyses that search for spectral hardening carried out by the IACT collaborations (who have access to the full instrumental response functions and raw data) would be extremely valuable.
This would enable a full likelihood analysis that could provide important 
insights not only into ALPs but also into other processes that might 
alter the $\gamma$-ray opacity such as electromagnetic 
cascades induced by ultra-high energy cosmic rays  \cite{essey2010b}
or the propagation of photons through cosmic voids \cite{furniss2014}.

\subsection{ALP-induced spectral irregularities}
\label{sec:irreg}

In case no spectral hardening is observed, 
it is difficult to constrain the photon-ALP coupling. 
The reason is that the non-observation might not be 
due to the absence of ALPs but due to a high-energy cutoff
of the intrinsic spectrum. 
This problem is alleviated when searching for  
spectral irregularities around $E_\mathrm{crit}$ 
and $E_\mathrm{max}$ as these features should 
be detectable as long as the energy resolution is sufficient to resolve the features and  that signal-to-noise ratio is 
high enough to distinguish the features from Poisson noise.\footnote{
Even for magnetic-field configurations in cell-like models that lead to no 
mixing in the strong mixing regime, the irregularities would still 
occur \cite{Meyer:2014mea}.
} 

Such analyses have been carried out at X-ray \cite{wouters2013,berg2016} and $\gamma$-ray energies \cite{hess2013:alps,ajello2016} with observations of central AGNs in galaxy clusters and groups.
 The authors employed more sophisticated 
$B$-field models in which the turbulent fields are described with 
suitable power spectra instead of cell-like morphologies. 
The non-observation of such features at X-rays leads 
to the exclusion of ALP masses $m_a < 10^{-11}$\,eV~\cite{wouters2013},
and $m_a < 10^{-12}$\,eV~\cite{berg2016} for $g_{a\gamma} \gtrsim 6\times10^{-12}\,\mathrm{GeV}^{-1}$.
These bounds also apply for lower masses since $\omega_\mathrm{pl} \gg m_a$, leading to irregularities independent of $m_a$.\footnote{ 
Interestingly, the conversions of ALPs from a 
cosmic ALP background could explain excess diffuse 
X-ray emission observed in several galaxy clusters \cite{Conlon:2013txa,Angus:2013sua}.}
At $\gamma$-ray energies, the strongest bounds on $g_{a\gamma}$ come from \emph{Fermi}-LAT observations of NGC\,1275
that are the most constraining limits between $0.5\,\mathrm{neV}\lesssim m_a\lesssim20\,\mathrm{neV}$ to date \cite{ajello2016}. 
Taken at face value, 
together with the bounds from the non-observation of irregularities in H.E.S.S. data of PKS\,2155-304 \cite{hess2013:alps} and the absence of a $\gamma$-ray burst signal
from SN\,1987A \cite{payez2014}, the possibility 
that ALPs alter the EBL $\gamma$-ray opacity is already seriously
constrained (see Fig. \ref{fig:rev}). 

\section{Conclusion and outlook}
\label{sec:out}
\begin{figure}
\centering
\includegraphics[width=.9\linewidth]{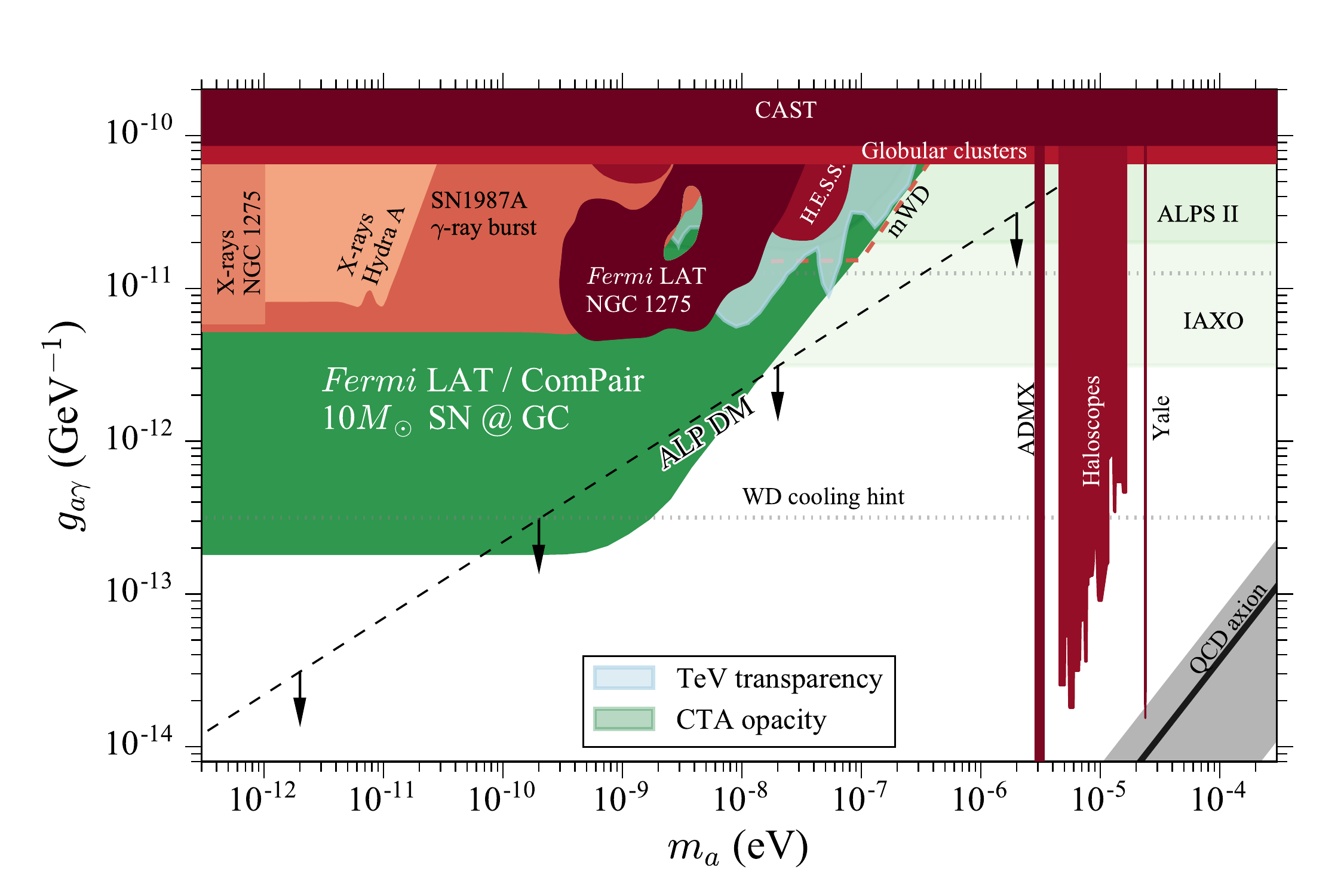}
\caption{The low-mass ALP parameter space. Current bounds are shown in red and include the non-observations of spectral irregularities at X-ray energies from 
Hydra\,A \cite{wouters2013} and NGC\,1275 \cite{berg2016}, at  $\gamma$-ray energies using the same source with the \emph{Fermi} LAT \cite{ajello2016} 
and PKS\,2155-304 with H.E.S.S. \cite{hess2013:alps}. 
Also shown are bounds from the CAST helioscope \cite{cast2007}, and haloscope microwave cavities looking for dark-matter axions and ALPs \cite{DePanfilis:1987dk,Wuensch:1989sa,Hagmann:1990tj,admx2010,Brubaker:2016ktl}, 
as well as limits inferred from globular cluster observations \cite{ayala2014}, and the non-detection of a $\gamma$-ray burst from SN\,1987A \cite{payez2014}.
Limits derived from optical polarization measurements (that could also be interpreted as a preferred region to explain such a polarization) of magnetic white dwarfs (mWD) are shown as a red dashed line \cite{gill2011}.
 ALP parameters that
could explain a low opacity of the Universe to $\gamma$ rays are shown in blue \cite{meyer2013} while those that would cause an additional white dwarf (WD) cooling lie between 
grey dashed lines \cite{isern2008}. 
Predicted parameters for one particular QCD axion model (the ``KVSZ'' axion~\cite{kim1979,vsz1980}) are shown together with an order of magnitude uncertainties (grey band).
Sensitivities of future experiments such as ALPS II \cite{alpsII}, IAXO \cite{irastorza2013} and 
future observations with CTA \cite{meyer2014cta} are shown in green along with the parameter space that could be probed with the \emph{Fermi} LAT in 
case of a core-collapse (10\,$M_\odot$) supernova (SN) in the Galactic center (GC)~\cite{Meyer:2016wrm}.
ALPs with parameters below the black dashed line could account for 
the entirety of the cold dark matter \cite{arias2012}.}
\label{fig:rev}
\end{figure}

Blazar observations at $\gamma$-ray energies
pose a complementary approach to search for ALPs, that 
could reveal themselves either through spectral irregularities 
or a boost of the $\gamma$-ray flux that would otherwise be attenuated 
through pair production. Current limits, future experimental sensitivities, and 
theoretically preferred regions for low mass ALPs are summarized in Fig. \ref{fig:rev}.

Future observations with CTA \cite{cta2011},
 HAWC \cite{hawc2013}, and HiSCORE \cite{hiscore2011}
with their good point source sensitivities also at energies beyond 1\,TeV
have the potential to search for ALPs at masses above 10\,neV.
Especially CTA will be able to probe the entire parameters space
where ALPs could explain the hints for a low $\gamma$-ray opacity \cite{meyer2014cta}.
The planned full-sky extragalactic CTA survey could also be used 
to search for a spectral hardening correlated with the photon-ALP conversion probability in Galactic magnetic fields \cite{wouters2014}.

ALPs could also be searched for with $\gamma$-ray observations using sources other than blazars.
Other possible sources to look for traces of ALPs include spectra of pulsars and observations of pulsar binary systems \cite{Dupays:2005xs}.
Another alternative is to search for a short $\gamma$-ray burst from the 
next Galactic core-collapse supernova. ALPs would be produced 
in such an event via the Primakoff process and subsequently convert 
into $\gamma$ rays in the Galactic magnetic field. 
These $\gamma$ rays would arrive simultaneously with the neutrinos 
produced in the supernova.
If such an event occurs during the lifetime of the \emph{Fermi} mission and
while the supernova is in the field of view of the satellite 
(the \emph{Fermi} LAT surveys $\sim20\,\%$ of the sky at any given 
moment), it would be possible to probe couplings down to
$2\times10^{-13}\,\mathrm{GeV}^{-1}$ for masses below 1\,neV \cite{Meyer:2016wrm}.  
Future $\gamma$-ray missions such as e-ASTROGAM~\cite{Tatischeff:2016ykb}, ComPair~\cite{Moiseev:2015lva},
or PANGU~\cite{Wu:2014tya} should even be more sensitive to such a signal 
given their accessible energy range and improved point spread function 
in comparison to the \emph{Fermi} LAT.

\section*{Acknowledgments}
MM is supported by a grant of the Knut and Alice Wallenberg Foundation, PI: Jan Conrad
 

\begin{footnotesize}

\begin{thebibliography}{100}

\bibitem{atwood2009}
W.~B. {Atwood}, A.~A. {Abdo}, M.~{Ackermann}, {\em et~al.}, ``{The Large Area
  Telescope on the Fermi Gamma-Ray Space Telescope Mission},'' {\em \apj},
  vol.~697, pp.~1071--1102, 2009.

\bibitem{hess}
F.~{Aharonian}, A.~G. {Akhperjanian}, A.~R. {Bazer-Bachi}, M.~{Beilicke},
  W.~{Benbow}, {\em et~al.}, ``{Observations of the Crab nebula with HESS},''
  {\em \aap}, vol.~457, pp.~899--915, Oct. 2006.

\bibitem{magic}
J.~Aleksić, S.~Ansoldi, L.~Antonelli, P.~Antoranz, A.~Babic, P.~Bangale,
  M.~Barceló, J.~Barrio, {\em et~al.}, ``The major upgrade of the magic
  telescopes, part ii: A performance study using observations of the crab
  nebula,'' {\em Astroparticle Physics}, vol.~72, pp.~76 -- 94, 2016.

\bibitem{veritas}
J.~{Holder}, V.~A. {Acciari}, E.~{Aliu}, T.~{Arlen}, M.~{Beilicke},
  W.~{Benbow}, {\em et~al.}, ``{Status of the VERITAS Observatory},'' in {\em
  American Institute of Physics Conference Series} (F.~A. {Aharonian},
  W.~{Hofmann}, and F.~{Rieger}, eds.), vol.~1085 of {\em American Institute of
  Physics Conference Series}, pp.~657--660, Dec. 2008.

\bibitem{Aharonian:2008kz}
F.~Aharonian {\em et~al.}, ``{Limits on an Energy Dependence of the Speed of
  Light from a Flare of the Active Galaxy PKS 2155-304},'' {\em Phys. Rev.
  Lett.}, vol.~101, p.~170402, 2008.

\bibitem{Ackermann:2009aa}
M.~Ackermann {\em et~al.}, ``{A limit on the variation of the speed of light
  arising from quantum gravity effects},'' {\em Nature}, vol.~462,
  pp.~331--334, 2009.

\bibitem{Ackermann:2015zua}
M.~Ackermann {\em et~al.}, ``{Searching for Dark Matter Annihilation from Milky
  Way Dwarf Spheroidal Galaxies with Six Years of Fermi Large Area Telescope
  Data},'' {\em Phys. Rev. Lett.}, vol.~115, no.~23, p.~231301, 2015.

\bibitem{hess:2016jja}
H.~Abdallah {\em et~al.}, ``{Search for dark matter annihilations towards the
  inner Galactic halo from 10 years of observations with H.E.S.S},'' {\em Phys.
  Rev. Lett.}, vol.~117, no.~11, p.~111301, 2016.

\bibitem{Zitzer:2015eqa}
B.~Zitzer, ``{A Search for Dark Matter from Dwarf Galaxies using VERITAS},''
  {\em PoS}, vol.~ICRC2015, p.~1225, 2016.

\bibitem{pq1977}
R.~D. {Peccei} and H.~R. {Quinn}, ``{CP conservation in the presence of
  pseudoparticles},'' {\em Physical Review Letters}, vol.~38, pp.~1440--1443,
  1977.

\bibitem{weinberg1978}
S.~{Weinberg}, ``{A new light boson?},'' {\em Physical Review Letters},
  vol.~40, pp.~223--226, 1978.

\bibitem{wilczek1978}
F.~{Wilczek}, ``{Problem of strong P and T invariance in the presence of
  instantons},'' {\em Physical Review Letters}, vol.~40, pp.~279--282, 1978.

\bibitem{jaeckel2010}
J.~{Jaeckel} and A.~{Ringwald}, ``{The Low-Energy Frontier of Particle
  Physics},'' {\em Annual Review of Nuclear and Particle Science}, vol.~60,
  pp.~405--437, 2010.

\bibitem{preskill1983}
J.~{Preskill}, M.~B. {Wise}, and F.~{Wilczek}, ``{Cosmology of the invisible
  axion},'' {\em Physics Letters B}, vol.~120, pp.~127--132, 1983.

\bibitem{abbott1983}
L.~F. {Abbott} and P.~{Sikivie}, ``{A cosmological bound on the invisible
  axion},'' {\em Physics Letters B}, vol.~120, pp.~133--136, Jan. 1983.

\bibitem{dine1983}
M.~{Dine} and W.~{Fischler}, ``{The not-so-harmless axion},'' {\em Physics
  Letters B}, vol.~120, pp.~137--141, Jan. 1983.

\bibitem{marsh2011}
D.~J.~E. {Marsh}, ``{Axiverse extended: Vacuum destabilization, early dark
  energy, and cosmological collapse},'' {\em \prd}, vol.~83, p.~123526, June
  2011.

\bibitem{arias2012}
P.~{Arias}, D.~{Cadamuro}, M.~{Goodsell}, J.~{Jaeckel}, J.~{Redondo}, and
  A.~{Ringwald}, ``{WISPy cold dark matter},'' {\em \jcap}, vol.~6, p.~13,
  2012.

\bibitem{sikivie1983}
P.~{Sikivie}, ``{Experimental tests of the 'invisible' axion},'' {\em Physical
  Review Letters}, vol.~51, pp.~1415--1417, Oct. 1983.

\bibitem{raffelt1988}
G.~{Raffelt} and L.~{Stodolsky}, ``{Mixing of the photon with low-mass
  particles},'' {\em \prd}, vol.~37, pp.~1237--1249, 1988.

\bibitem{witten1984}
E.~{Witten}, ``{Some properties of O(32) superstrings},'' {\em Physics Letters
  B}, vol.~149, pp.~351--356, 1984.

\bibitem{arvanitaki2010}
A.~{Arvanitaki}, S.~{Dimopoulos}, S.~{Dubovsky}, {\em et~al.}, ``{String
  axiverse},'' {\em \prd}, vol.~81, p.~123530, June 2010.

\bibitem{ringwald2014review}
A.~{Ringwald}, ``{Axions and Axion-Like Particles},'' {\em ArXiv e-prints},
  July 2014.

\bibitem{Giannotti:2010ty}
M.~Giannotti, L.~D. Duffy, and R.~Nita, ``{New constraints for heavy axion-like
  particles from supernovae},'' {\em JCAP}, vol.~1101, p.~015, 2011.

\bibitem{Berenji:2016jji}
B.~Berenji, J.~Gaskins, and M.~Meyer, ``{Constraints on Axions and Axionlike
  Particles from Fermi Large Area Telescope Observations of Neutron Stars},''
  {\em Phys. Rev.}, vol.~D93, no.~4, p.~045019, 2016.

\bibitem{dobrynina2015}
A.~{Dobrynina}, A.~{Kartavtsev}, and G.~{Raffelt}, ``{Photon-photon dispersion
  of TeV gamma rays and its role for photon-ALP conversion},'' {\em \prd},
  vol.~91, p.~083003, Apr. 2015.

\bibitem{grossmann2002}
Y.~{Grossman}, S.~{Roy}, and J.~{Zupan}, ``{Effects of initial axion production
  and photon-axion oscillation on type Ia supernova dimming [rapid
  communication]},'' {\em Physics Letters B}, vol.~543, pp.~23--28, 2002.

\bibitem{csaki2003}
C.~{Cs{\'a}ki}, N.~{Kaloper}, M.~{Peloso}, and J.~{Terning}, ``{Super-GZK
  photons from photon axion mixing},'' {\em \jcap}, vol.~5, p.~5, 2003.

\bibitem{mirizzi2008}
A.~{Mirizzi}, G.~G. {Raffelt}, and P.~D. {Serpico}, ``{Photon-Axion Conversion
  in Intergalactic Magnetic Fields and Cosmological Consequences},'' in {\em
  Axions} (M.~{Kuster}, G.~{Raffelt}, and B.~{Beltr{\'a}n}, eds.), vol.~741 of
  {\em Lecture Notes in Physics, Berlin Springer Verlag}, p.~115, 2008.

\bibitem{bassan2010}
N.~{Bassan}, A.~{Mirizzi}, and M.~{Roncadelli}, ``{Axion-like particle effects
  on the polarization of cosmic high-energy gamma sources},'' {\em \jcap},
  vol.~5, p.~10, 2010.

\bibitem{deangelis2011}
A.~{de Angelis}, G.~{Galanti}, and M.~{Roncadelli}, ``{Relevance of axionlike
  particles for very-high-energy astrophysics},'' {\em \prd}, vol.~84, no.~10,
  p.~105030, 2011.

\bibitem{meyer2014}
M.~{Meyer}, D.~{Montanino}, and J.~{Conrad}, ``{On detecting oscillations of
  gamma rays into axion-like particles in turbulent and coherent magnetic
  fields},'' {\em \jcap}, vol.~9, p.~3, 2014.

\bibitem{urry1995}
C.~M. {Urry} and P.~{Padovani}, ``{Unified Schemes for Radio-Loud Active
  Galactic Nuclei},'' {\em \pasp}, vol.~107, pp.~803--+, 1995.

\bibitem{3fgl}
M.~Ackermann {\em et~al.}, ``{The Third Catalog of Active Galactic Nuclei
  Detected by the Fermi Large Area Telescope},'' {\em Astrophys. J.}, vol.~810,
  no.~1, p.~14, 2015.

\bibitem{nikishov1962}
A.~I. {Nikishov}, ``{Absorption of high-energy photons in the Universe},'' {\em
  Sov. Phys. JETP}, vol.~14, pp.~393--394, 1962.

\bibitem{gould1967}
R.~J. {Gould} and G.~P. {Schr{\'e}der}, ``{Pair Production in Photon-Photon
  Collisions},'' {\em Physical Review}, vol.~155, pp.~1404--1407, 1967.

\bibitem{gould1967a}
R.~J. {Gould} and G.~P. {Schr{\'e}der}, ``{Opacity of the Universe to
  High-Energy Photons},'' {\em Physical Review}, vol.~155, pp.~1408--1411, Mar.
  1967.

\bibitem{sanchezconde2009}
M.~A. {S{\'a}nchez-Conde}, D.~{Paneque}, E.~{Bloom}, {\em et~al.}, ``{Hints of
  the existence of axionlike particles from the gamma-ray spectra of
  cosmological sources},'' {\em \prd}, vol.~79, no.~12, p.~123511, 2009.

\bibitem{tavecchio2012}
F.~{Tavecchio}, M.~{Roncadelli}, G.~{Galanti}, and G.~{Bonnoli}, ``{Evidence
  for an axion-like particle from PKS 1222+216?},'' {\em \prd}, vol.~86, no.~8,
  p.~085036, 2012.

\bibitem{mena2013}
O.~{Mena} and S.~{Razzaque}, ``{Hints of an axion-like particle mixing in the
  GeV gamma-ray blazar data?},'' {\em \jcap}, vol.~11, p.~23, 2013.

\bibitem{tavecchio2014}
F.~{Tavecchio}, M.~{Roncadelli}, and G.~{Galanti}, ``{Photons into axion-like
  particles conversion in Active Galactic Nuclei},'' {\em ArXiv e-prints},
  2014.

\bibitem{meyer2014cta}
M.~{Meyer} and J.~{Conrad}, ``{Sensitivity of the Cherenkov Telescope Array to
  the detection of axion-like particles at high gamma-ray opacities},'' {\em
  \jcap}, vol.~12, p.~16, 2014.

\bibitem{horns2012ICM}
D.~{Horns}, L.~{Maccione}, M.~{Meyer}, A.~{Mirizzi}, D.~{Montanino}, and
  M.~{Roncadelli}, ``{Hardening of TeV gamma spectrum of active galactic nuclei
  in galaxy clusters by conversions of photons into axionlike particles},''
  {\em \prd}, vol.~86, no.~7, p.~075024, 2012.

\bibitem{hess2013:alps}
A.~{Abramowski}, F.~{Acero}, F.~{Aharonian}, {\em et~al.}, ``{Constraints on
  axionlike particles with H.E.S.S. from the irregularity of the PKS 2155-304
  energy spectrum},'' {\em \prd}, vol.~88, no.~10, p.~102003, 2013.

\bibitem{Kartavtsev:2016doq}
A.~Kartavtsev, G.~Raffelt, and H.~Vogel, ``{Extragalactic photon-ALP conversion
  at CTA energies},'' 2016.

\bibitem{deangelis2007}
A.~{de Angelis}, M.~{Roncadelli}, and O.~{Mansutti}, ``{Evidence for a new
  light spin-zero boson from cosmological gamma-ray propagation?},'' {\em
  \prd}, vol.~76, no.~12, p.~121301, 2007.

\bibitem{mirizzi2007}
A.~{Mirizzi}, G.~G. {Raffelt}, and P.~D. {Serpico}, ``{Signatures of axionlike
  particles in the spectra of TeV gamma-ray sources},'' {\em \prd}, vol.~76,
  no.~2, p.~023001, 2007.

\bibitem{simet2008}
M.~{Simet}, D.~{Hooper}, and P.~D. {Serpico}, ``{Milky Way as a
  kiloparsec-scale axionscope},'' {\em \prd}, vol.~77, no.~6, p.~063001, 2008.

\bibitem{mirizzi2009}
A.~{Mirizzi} and D.~{Montanino}, ``{Stochastic conversions of TeV photons into
  axion-like particles in extragalactic magnetic fields},'' {\em \jcap},
  vol.~12, p.~4, 2009.

\bibitem{pshirkov2016}
M.~S. {Pshirkov}, P.~G. {Tinyakov}, and F.~R. {Urban}, ``{New Limits on
  Extragalactic Magnetic Fields from Rotation Measures},'' {\em Physical Review
  Letters}, vol.~116, p.~191302, May 2016.

\bibitem{pshirkov2011}
M.~S. {Pshirkov}, P.~G. {Tinyakov}, P.~P. {Kronberg}, and K.~J. {Newton-McGee},
  ``{Deriving the Global Structure of the Galactic Magnetic Field from Faraday
  Rotation Measures of Extragalactic Sources},'' {\em \apj}, vol.~738, p.~192,
  2011.

\bibitem{jansson2012}
R.~{Jansson} and G.~R. {Farrar}, ``{A New Model of the Galactic Magnetic
  Field},'' {\em \apj}, vol.~757, p.~14, 2012.

\bibitem{adam2016}
R.~Adam {\em et~al.}, ``{Planck intermediate results. XLII. Large-scale
  Galactic magnetic fields},'' 2016.

\bibitem{dwek2013}
E.~{Dwek} and F.~{Krennrich}, ``{The extragalactic background light and the
  gamma-ray opacity of the universe},'' {\em Astroparticle Physics}, vol.~43,
  pp.~112--133, 2013.

\bibitem{hauser1998}
M.~G. {Hauser}, R.~G. {Arendt}, {\em et~al.}, ``{The COBE Diffuse Infrared
  Background Experiment Search for the Cosmic Infrared Background. I. Limits
  and Detections},'' {\em \apj}, vol.~508, pp.~25--43, 1998.

\bibitem{hauser2001}
M.~G. {Hauser} and E.~{Dwek}, ``{The Cosmic Infrared Background: Measurements
  and Implications},'' {\em \araa}, vol.~39, pp.~249--307, 2001.

\bibitem{kashlinsky2005}
A.~{Kashlinsky}, ``{Cosmic infrared background and early galaxy evolution
  [review article]},'' {\em \physrep}, vol.~409, pp.~361--438, 2005.

\bibitem{franceschini2008}
A.~{Franceschini}, G.~{Rodighiero}, and M.~{Vaccari}, ``{Extragalactic
  optical-infrared background radiation, its time evolution and the cosmic
  photon-photon opacity},'' {\em \aap}, vol.~487, pp.~837--852, 2008.

\bibitem{finke2010}
J.~D. {Finke}, S.~{Razzaque}, and C.~D. {Dermer}, ``{Modeling the Extragalactic
  Background Light from Stars and Dust},'' {\em \apj}, vol.~712, pp.~238--249,
  Mar. 2010.

\bibitem{kneiske2010}
T.~M. {Kneiske} and H.~{Dole}, ``{A lower-limit flux for the extragalactic
  background light},'' {\em \aap}, vol.~515, pp.~A19+, 2010.

\bibitem{dominguez2011}
A.~{Dom{\'{\i}}nguez}, J.~R. {Primack}, D.~J. {Rosario}, {\em et~al.},
  ``{Extragalactic background light inferred from AEGIS galaxy-SED-type
  fractions},'' {\em \mnras}, vol.~410, pp.~2556--2578, 2011.

\bibitem{gilmore2012}
R.~C. {Gilmore}, R.~S. {Somerville}, J.~R. {Primack}, and
  A.~{Dom{\'{\i}}nguez}, ``{Semi-analytic modelling of the extragalactic
  background light and consequences for extragalactic gamma-ray spectra},''
  {\em \mnras}, vol.~422, pp.~3189--3207, 2012.

\bibitem{inoue2013}
Y.~{Inoue}, S.~{Inoue}, M.~A.~R. {Kobayashi}, R.~{Makiya}, Y.~{Niino}, and
  T.~{Totani}, ``{Extragalactic Background Light from Hierarchical Galaxy
  Formation: Gamma-Ray Attenuation up to the Epoch of Cosmic Reionization and
  the First Stars},'' {\em \apj}, vol.~768, p.~197, May 2013.

\bibitem{deangelis2009}
A.~{De Angelis}, O.~{Mansutti}, M.~{Persic}, and M.~{Roncadelli}, ``{Photon
  propagation and the very high energy {$\gamma$}-ray spectra of blazars: how
  transparent is the Universe?},'' {\em \mnras}, vol.~394, pp.~L21--L25, 2009.

\bibitem{dominguez2011alps}
A.~{Dom{\'{\i}}nguez}, M.~A. {S{\'a}nchez-Conde}, and F.~{Prada}, ``{Axion-like
  particle imprint in cosmological very-high-energy sources},'' {\em \jcap},
  vol.~11, p.~20, 2011.

\bibitem{essey2012}
W.~{Essey} and A.~{Kusenko}, ``{On Weak Redshift Dependence of Gamma-Ray
  Spectra of Distant Blazars},'' {\em \apjl}, vol.~751, p.~L11, 2012.

\bibitem{rubtsov2014}
G.~Rubtsov and S.~Troitsky, ``{Breaks in gamma-ray spectra of distant blazars
  and transparency of the Universe},'' {\em JETP Lett.}, vol.~100, no.~6,
  pp.~397--401, 2014.

\bibitem{galanti2015}
G.~{Galanti}, M.~{Roncadelli}, A.~{De Angelis}, and G.~F. {Bignami},
  ``{Axion-like particles explain the unphysical redshift-dependence of AGN
  gamma-ray spectra},'' {\em ArXiv e-prints}, 2015.

\bibitem{horns2012}
D.~{Horns} and M.~{Meyer}, ``{Indications for a pair-production anomaly from
  the propagation of VHE gamma-rays},'' {\em \jcap}, vol.~2, p.~33, 2012.

\bibitem{meyer2012ppa}
M.~{Meyer}, D.~{Horns}, and M.~{Raue}, ``{Revisiting the Indication for a low
  opacity Universe for very high energy gamma-rays},'' {\em ArXiv e-prints},
  Nov. 2012.

\bibitem{meyer2013}
M.~{Meyer}, D.~{Horns}, and M.~{Raue}, ``{First lower limits on the
  photon-axion-like particle coupling from very high energy gamma-ray
  observations},'' {\em \prd}, vol.~87, no.~3, p.~035027, 2013.

\bibitem{lefa2011b}
E.~{Lefa}, F.~A. {Aharonian}, and F.~M. {Rieger}, ``{''Leading Blob'' Model in
  a Stochastic Acceleration Scenario: The Case of the 2009 Flare of Mkn 501},''
  {\em \apjl}, vol.~743, p.~L19, 2011.

\bibitem{biteau2015}
J.~{Biteau} and D.~A. {Williams}, ``{The Extragalactic Background Light, the
  Hubble Constant, and Anomalies: Conclusions from 20 Years of TeV Gamma-Ray
  Observations},'' {\em \apj}, vol.~812, p.~60, 2015.

\bibitem{Sanchez:2013lla}
D.~A. Sanchez, S.~Fegan, and B.~Giebels, ``{Evidence for a cosmological effect
  in gamma-ray spectra of BL Lacs},'' {\em Astron. Astrophys.}, vol.~554,
  p.~A75, 2013.

\bibitem{dominguez2015}
A.~{Dom{\'{\i}}nguez} and M.~{Ajello}, ``{Spectral Analysis of Fermi-LAT
  Blazars above 50 GeV},'' {\em \apjl}, vol.~813, p.~L34, 2015.

\bibitem{2fhl}
M.~{Ackermann} {\em et~al.}, ``{2FHL: The Second Catalog of Hard Fermi-LAT
  Sources},'' {\em \apjs}, vol.~222, p.~5, 2016.

\bibitem{essey2010b}
W.~{Essey}, O.~E. {Kalashev}, A.~{Kusenko}, and J.~F. {Beacom}, ``{Secondary
  Photons and Neutrinos from Cosmic Rays Produced by Distant Blazars},'' {\em
  Physical Review Letters}, vol.~104, no.~14, p.~141102, 2010.

\bibitem{furniss2014}
A.~{Furniss}, P.~M. {Sutter}, J.~R. {Primack}, and A.~{Dominguez}, ``{A
  Correlation Between Hard Gamma-ray Sources and Cosmic Voids Along the Line of
  Sight},'' {\em ArXiv e-prints}, 2014.

\bibitem{Meyer:2014mea}
M.~Meyer, ``{Modelling gamma-ray-axion-like particle oscillations in turbulent
  magnetic fields: relevance for observations with Cherenkov telescopes},'' in
  {\em {Proceedings, 10th Patras Workshop on Axions, WIMPs and WISPs
  (AXION-WIMP 2014): Geneva, Switzerland, June 29-July 4, 2014}}, 2014.

\bibitem{wouters2013}
D.~{Wouters} and P.~{Brun}, ``{Constraints on Axion-like Particles from X-Ray
  Observations of the Hydra Galaxy Cluster},'' {\em \apj}, vol.~772, p.~44,
  July 2013.

\bibitem{berg2016}
M.~Berg, J.~P. Conlon, F.~Day, N.~Jennings, S.~Krippendorf, A.~J. Powell, and
  M.~Rummel, ``{Searches for Axion-Like Particles with NGC1275: Observation of
  Spectral Modulations},'' 2016.

\bibitem{ajello2016}
M.~Ajello {\em et~al.}, ``{Search for Spectral Irregularities due to
  Photon–Axionlike-Particle Oscillations with the Fermi Large Area
  Telescope},'' {\em Phys. Rev. Lett.}, vol.~116, no.~16, p.~161101, 2016.

\bibitem{Conlon:2013txa}
J.~P. Conlon and M.~C.~D. Marsh, ``{Excess Astrophysical Photons from a 0.1–1
  keV Cosmic Axion Background},'' {\em Phys. Rev. Lett.}, vol.~111, no.~15,
  p.~151301, 2013.

\bibitem{Angus:2013sua}
S.~Angus, J.~P. Conlon, M.~C.~D. Marsh, A.~J. Powell, and L.~T. Witkowski,
  ``{Soft X-ray Excess in the Coma Cluster from a Cosmic Axion Background},''
  {\em JCAP}, vol.~1409, no.~09, p.~026, 2014.

\bibitem{payez2014}
A.~Payez, C.~Evoli, T.~Fischer, M.~Giannotti, A.~Mirizzi, and A.~Ringwald,
  ``{Revisiting the SN1987A gamma-ray limit on ultralight axion-like
  particles},'' {\em JCAP}, vol.~1502, no.~02, p.~006, 2015.

\bibitem{cta2011}
M.~{Actis}, G.~{Agnetta}, F.~{Aharonian}, {\em et~al.}, ``{Design concepts for
  the Cherenkov Telescope Array CTA: an advanced facility for ground-based
  high-energy gamma-ray astronomy},'' {\em Experimental Astronomy}, vol.~32,
  pp.~193--316, 2011.

\bibitem{hawc2013}
A.~U. {Abeysekara}, R.~{Alfaro}, C.~{Alvarez}, {\em et~al.}, ``{Sensitivity of
  the high altitude water Cherenkov detector to sources of multi-TeV gamma
  rays},'' {\em Astroparticle Physics}, vol.~50, pp.~26--32, 2013.

\bibitem{hiscore2011}
M.~{Tluczykont}, D.~{Hampf}, D.~{Horns}, {\em et~al.}, ``{The ground-based
  large-area wide-angle {$\gamma$}-ray and cosmic-ray experiment HiSCORE},''
  {\em Advances in Space Research}, vol.~48, pp.~1935--1941, 2011.

\bibitem{wouters2014}
D.~{Wouters} and P.~{Brun}, ``{Anisotropy test of the axion-like particle
  Universe opacity effect: a case for the Cherenkov Telescope Array},'' {\em
  \jcap}, vol.~1, p.~16, 2014.

\bibitem{Dupays:2005xs}
A.~Dupays, C.~Rizzo, M.~Roncadelli, and G.~F. Bignami, ``{Looking for light
  pseudoscalar bosons in the binary pulsar system j0737-3039},'' {\em Phys.
  Rev. Lett.}, vol.~95, p.~211302, 2005.

\bibitem{Meyer:2016wrm}
M.~Meyer, M.~Giannotti, A.~Mirizzi, J.~Conrad, and M.~Sanchez-Conde, ``{The
  Fermi Large Area Telescope as a Galactic Supernovae Axionscope},'' 2016.

\bibitem{Tatischeff:2016ykb}
V.~Tatischeff {\em et~al.}, ``{The e-ASTROGAM gamma-ray space mission},'' {\em
  Proc. SPIE Int. Soc. Opt. Eng.}, vol.~9905, p.~99052N, 2016.

\bibitem{Moiseev:2015lva}
A.~A. Moiseev {\em et~al.}, ``{Compton-Pair Production Space Telescope
  (ComPair) for MeV Gamma-ray Astronomy},'' 2015.

\bibitem{Wu:2014tya}
X.~Wu, M.~Su, A.~Bravar, J.~Chang, Y.~Fan, M.~Pohl, and R.~Walter, ``{PANGU: A
  High Resolution Gamma-ray Space Telescope},'' {\em Proc. SPIE Int. Soc. Opt.
  Eng.}, vol.~9144, p.~91440F, 2014.

\bibitem{cast2007}
S.~{Andriamonje}, S.~{Aune}, D.~{Autiero}, {CAST Collaboration}, {\em et~al.},
  ``{An improved limit on the axion photon coupling from the CAST
  experiment},'' {\em \jcap}, vol.~4, p.~10, 2007.

\bibitem{DePanfilis:1987dk}
S.~De~Panfilis, A.~C. Melissinos, B.~E. Moskowitz, J.~T. Rogers, Y.~K.
  Semertzidis, W.~Wuensch, H.~J. Halama, A.~G. Prodell, W.~B. Fowler, and F.~A.
  Nezrick, ``{Limits on the Abundance and Coupling of Cosmic Axions at
  4.5-Microev < m(a) < 5.0-Microev},'' {\em Phys. Rev. Lett.}, vol.~59, p.~839,
  1987.

\bibitem{Wuensch:1989sa}
W.~Wuensch, S.~De~Panfilis-Wuensch, Y.~K. Semertzidis, J.~T. Rogers, A.~C.
  Melissinos, H.~J. Halama, B.~E. Moskowitz, A.~G. Prodell, W.~B. Fowler, and
  F.~A. Nezrick, ``{Results of a Laboratory Search for Cosmic Axions and Other
  Weakly Coupled Light Particles},'' {\em Phys. Rev.}, vol.~D40, p.~3153, 1989.

\bibitem{Hagmann:1990tj}
C.~Hagmann, P.~Sikivie, N.~S. Sullivan, and D.~B. Tanner, ``{Results from a
  search for cosmic axions},'' {\em Phys. Rev.}, vol.~D42, pp.~1297--1300,
  1990.

\bibitem{admx2010}
S.~J. Asztalos, G.~Carosi, C.~Hagmann, D.~Kinion, K.~van Bibber, M.~Hotz, L.~J.
  Rosenberg, G.~Rybka, J.~Hoskins, J.~Hwang, P.~Sikivie, D.~B. Tanner,
  R.~Bradley, and J.~Clarke, ``Squid-based microwave cavity search for
  dark-matter axions,'' {\em Phys. Rev. Lett.}, vol.~104, p.~041301, Jan 2010.

\bibitem{Brubaker:2016ktl}
B.~M. Brubaker {\em et~al.}, ``{First results from a microwave cavity axion
  search at 24 micro-eV},'' 2016.

\bibitem{ayala2014}
A.~Ayala, I.~{Dom{\'{\i}}nguez}, M.~Giannotti, {\em et~al.}, ``{Revisiting the
  bound on axion-photon coupling from Globular Clusters},'' {\em Phys. Rev.
  Lett.}, vol.~113, no.~19, p.~191302, 2014.

\bibitem{gill2011}
R.~{Gill} and J.~S. {Heyl}, ``{Constraining the photon-axion coupling constant
  with magnetic white dwarfs},'' {\em \prd}, vol.~84, no.~8, p.~085001, 2011.

\bibitem{isern2008}
J.~{Isern}, E.~{Garc{\'{\i}}a-Berro}, S.~{Torres}, and S.~{Catal{\'a}n},
  ``{Axions and the Cooling of White Dwarf Stars},'' {\em \apjl}, vol.~682,
  pp.~L109--L112, 2008.

\bibitem{kim1979}
J.~E. Kim, ``{Weak Interaction Singlet and Strong CP Invariance},'' {\em Phys.
  Rev. Lett.}, vol.~43, p.~103, 1979.

\bibitem{vsz1980}
M.~A. Shifman, A.~I. Vainshtein, and V.~I. Zakharov, ``{Can Confinement Ensure
  Natural CP Invariance of Strong Interactions?},'' {\em Nucl. Phys.},
  vol.~B166, pp.~493--506, 1980.

\bibitem{alpsII}
R.~{B{\"a}hre}, B.~{D{\"o}brich}, J.~{Dreyling-Eschweiler}, {\em et~al.},
  ``{Any light particle search II {--} Technical Design Report},'' {\em Journal
  of Instrumentation}, vol.~8, p.~9001, 2013.

\bibitem{irastorza2013}
I.~G. {Irastorza}, F.~T. {Avignone}, G.~{Cantatore}, {\em et~al.}, ``{Future
  axion searches with the International Axion Observatory (IAXO)},'' {\em
  Journal of Physics Conference Series}, vol.~460, no.~1, p.~012002, 2013.

\end{thebibliography}

\end{footnotesize}


\end{document}